\documentclass[manuscript,screen]{acmart}

\usepackage{arxiv26-uniqueness-aware-feature-engineering-frame}

\begin{document}
\title{Feature Selection of Human Classification Targets under Uniqueness}
\title{Representation Collisions between Two Representations of the Same Instances}
\title{Making use of high dimensional data despite the curse of dimensionality}
\title{Collision Probabilities for pairs of representations of the same human decision subject}
\title{How can we select features that Represent Human Decision Targets Appropriately?}
\title{Human Feature Selection as a Combinatorial Problem}
\title{A probabilistic framework for computing representation fidelity}
\title{Defying the odds in algorithmic decision making about humans}
\title{Human-Centric Feature Selection: Representing Individuals for Algorithmic Decision Making}
\title{Uniqueness-Sensitive Feature Selection for Human Decision Targets}
\title{Uniqueness-Aware Feature Selection for Representing Humans}
\title{Uniqueness-Aware Feature Engineering for Representing Humans in Algorithmic Decision Making}

\author{Theresa Elstner}
\email{theresa.elstner@uni-kassel.de}
\orcid{0000-0002-6359-0752}
\affiliation{%
  \institution{Kassel University}
  \city{Kassel}
  \country{Germany}
}

\author{Martin Potthast}
\email{martin.potthast@uni-kassel.de}
\orcid{0000-0003-2451-0665}
\affiliation{%
  \institution{Kassel University}
  \city{Kassel}
  \country{Germany}
}

\begin{abstract}
Fairness analyses of algorithmic decision systems typically treat the feature representation of individuals as an a priori fixed input. How well a feature representation captures what is actually decision-relevant for different people and their unique backgrounds, however, is typically left unexamined. We address this gap by modelling feature engineering as a stochastic selection process over the pooled characteristics as candidate features for a finite training population. Based on this model, we develop a new framework for uniqueness-aware feature engineering which operates in two layers. The first layer measures population heterogeneity using a multivariate hypergeometric model of feature prevalence. It shows how well an individual is represented by any given input representation and which individual of the poplutation is ``least represented'' compared to their personal, reference characteristics. The second layer acts on this assessment using set-similarity measures to search for an optimum between representational fit of the population and predictive accuracy. Without assuming prior knowledge about unseen individuals, our framework grants represented humans agency over what counts as decision-relevant in their case, gives deploying institutions accountability through a publishable representation fit report, and provides each person subject to a decision transparency about their personal representation fit. We operationalize our framework as a six-step pipeline and show a proof-of-principle using synthetic data. Our approach offers a blueprint for making algorithmic decision systems accountable for how well they represent individuals during development and in production.
\end{abstract}



\setcopyright{acmlicensed}
\copyrightyear{2018}
\acmYear{2018}
\acmDOI{XXXXXXX.XXXXXXX}
\acmConference[Conference acronym 'XX]{Make sure to enter the correct
  conference title from your rights confirmation email}{June 03--05,
  2018}{Woodstock, NY}
\acmISBN{978-1-4503-XXXX-X/2018/06}

\setcopyright{none}



\maketitle

\section{Introduction}

\begin{figure}[h]
\centering
\includegraphics[width=\linewidth]{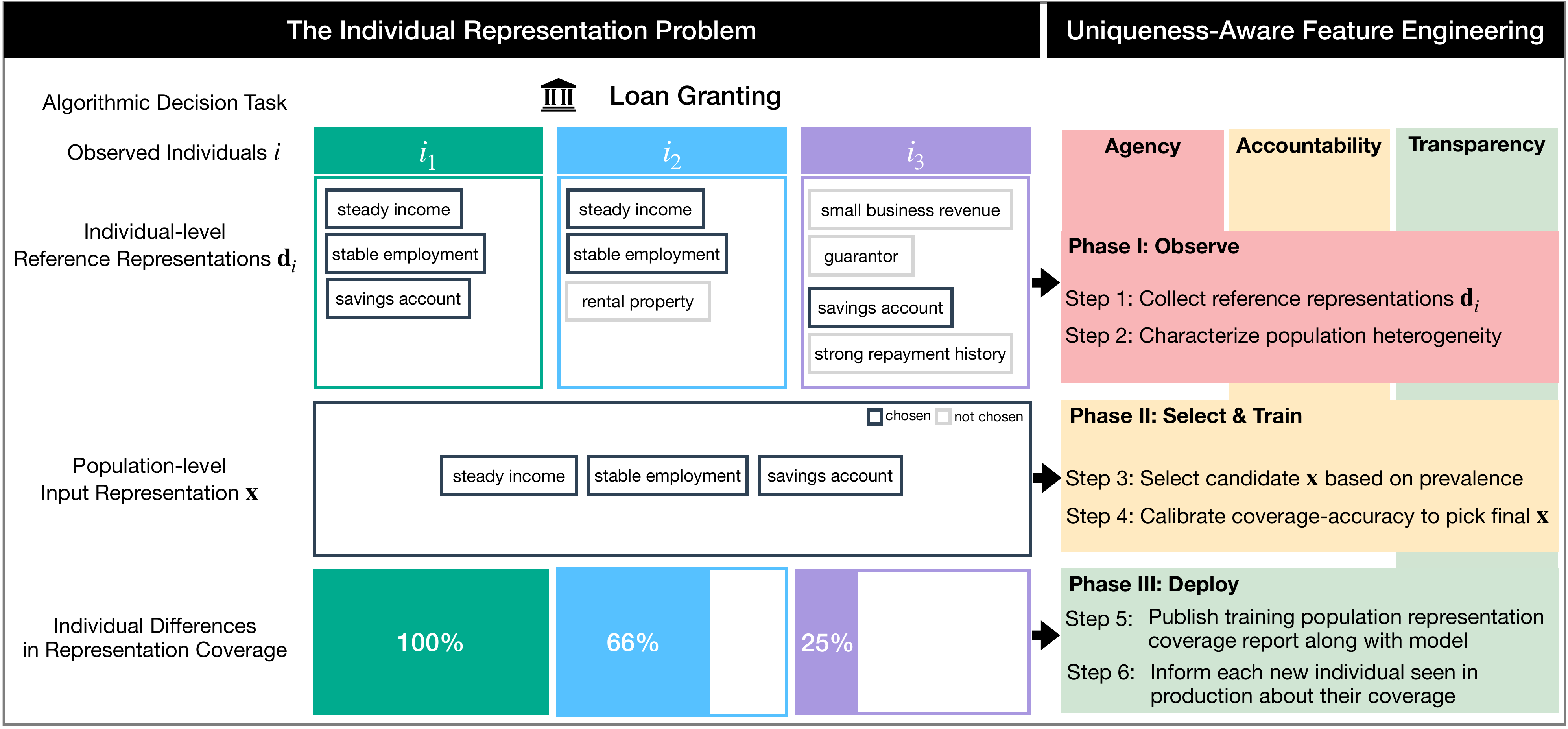}
\caption{\emph{Left:} The individual representation problem: individuals $i_1$ to $i_3$ each hold a unique set of task-relevant features (their reference representations $\mathbf{d_i}$) A single fixed input representation $\mathbf{x}$ grounded in the population covers individual reference representations to different degrees: It captures many features of individual $i_1$ and only some of $i_2$ and $i_3$. \emph{Right:} The proposed six-step pipeline for uniqueness-aware feature engineering, organized in three phases: (I) Observe---collect reference representations and characterize group heterogeneity provides humans with agency towards their representation, (II) Select and Train---choose $\mathbf{x}$ from the observed population and train maximizing representation coverage and prediction accuracy gives deploying institutions a tool to account for their choice of input representation, and (III) Deploy---publish a representation fit report for the observed population and inform individuals in production about their coverage to uphold transparency over time.}
\label{fig:uniqueness-aware-framework}
\end{figure}

Eightynine-year-olds need walking aids. High-school dropouts won't earn much later in their career. Disinterested first-year students won't graduate with a~PhD. Such and similar observations can plausibly result from group-level data analyses, but individual observations can reverse each conclusion: an 89-year-old is an active gymnast, a dropout built a successful company, and an unmotivated student was just lacking a challenge.

Group-level representations like these lead to unjustified and incorrect decisions for individuals, i.e., decisions based on incomplete knowledge or ill-fitting observations.%
\footnote{Note that even correct decisions can be unjustified when they are made for the wrong reasons.}
Algorithmic decision about humans, such as loan granting or college admission, is particularly prone to this problem: they are built on the assumption that a single, fixed input representation captures all humans equally well. We call this the \emph{individual representation problem} and illustrate it in~Figure~\ref{fig:uniqueness-aware-framework}~(left): as individuals differ in which characteristics are genuinely predictive for their case, no single fixed input representation can faithfully capture an a heterogeneous population. This creates an inherent tension between the breadth of representational coverage and the accuracy of decision outcomes.

In this work, we ask ``How can we design feature engineering strategies that account for human uniqueness and group heterogeneity?'' We inspect feature engineering strategies that yield representations of humans used as input to algorithmic decision making systems (we call them \emph{input representations}). In contrast to traditional feature engineering strategies, uniqueness-aware feature engineering takes into account unique circumstances and individual characteristics by gathering individual \emph{reference representations}, i.e. representations of humans that document task-relevant features from their perspective. Combined with a set of measures that assess the validity, i.e., the \emph{fit}, of the input representation with respect to group heterogeneity, and similarity to reference representations, uniqueness-aware feature engineering provides a framework for affected individuals to detect and contest unjustified decisions, as well as for developers to optimize representation fit and decision accuracy.

We address the individual representation problem with our uniqueness-aware feature engineering pipeline illustrated in Figure~\ref{fig:uniqueness-aware-framework}. It entails six steps:
\Ni
collecting reference representation data that reflects the internal perspective of individuals on what is decision-relevant for them;
\Nii
characterizing the heterogeneity of the observed group using the measures we introduce; 
\Niii
selecting from the most prevalent features in the pooled population while excluding features too rare to support reliable decision making;
\Niv
calibrating the optimum between coverage and accuracy explicitly rather than absorbing it silently into model training;
\Nv
publishing any deployed model together with a structured account of the individual representation problem; and
\Nvi
in production, informing each new individual of how well the chosen input representation fits their specific reference representation.

The pipeline is organized in three phases; each addresses a core desideratum of trustworthy algorithmic decision-making by design: Collecting reference representations grants individuals agency over what counts as decision-relevant in their case. Selecting and training with explicit coverage calibration gives deploying institutions a quantitative basis on which to be held accountable for their choice of input representation, and publishing the representation fit report alongside informing individuals at deployment upholds transparency over time.

In summary, our contributions are
\Ni
a framework for uniqueness-aware feature engineering for representing human individuals (cf.\ Figure~\ref{fig:uniqueness-aware-framework}),
\Nii
a set of measures to quantify the faithfulness of each individual representation,
\Niii
a proof-of-principle analysis on synthetic data illustrating the framework's potential, and
\Niv
a discussion of practical implications for feature engineering processes and auditing of algorithmic decisions with human decision targets.

\section{Related Work}
\label{related-work}

Algorithmic decisions about people require abstract representations of people. These representations reduce individuals to a set of observable features and thereby necessarily omit aspects of their unique reality. The central question is therefore not only whether a feature set enables a useful prediction, but whether it adequately represents the people to whom the resulting decisions apply.

\paragraph{Feature engineering, feature selection, and representation}
Feature engineering converts unstructured and often high-dimensional raw data to computationally processable features that are more structured and often of lower dimensionality. \citet{verdonck:2024} survey and classify various feature engineering strategies across domains, differentiating two major strategies: adjusting existing features using transformations or creating new features. Traditional feature engineering strategies include dimensionality reduction techniques, such as PCA. Reducing raw data to the most frequent and predictive features allows computation of more accurate results over the entire population but disregards individual features that only apply to a few, reinforcing the assumption that those reductions have the same effect on all individuals.

A structured account of this problem is offered by \citet{friedler:2021}, who distinguish between idealized \emph{construct spaces} for features and decisions and their imperfect observed counterparts used as proxies in formal decision systems. \citet{jacobs:2021} refine this further through measurement modeling, showing that construct--proxy mismatches lead to systematic errors and inequities when observable features fail to operationalize task-relevant constructs. \citet{suresh:2021} identify three sources of such measurement bias: construct oversimplification, variation in feature measurement across prediction targets, and variation in measurement accuracy across groups. Taken together, these works establish that feature selection is not a neutral technical choice but a locus of potential harm.

\citet{grgic-hlaca:2018} show that people assess features as unfair when they are perceived as irrelevant or unreliable with respect to the decision at hand. This highlights the role of perceived relevance in decision justification and motivates approaches that measure misalignment between individuals' relevant traits and the features used by the algorithm. Work considering fair use of features from the perspective of represented individuals shows that each person would select different sets to represent them, even within a closed set~\cite{grgic-hlaca:2018}. However, these approaches still operate within a predefined feature space that hinders representation of unique entities such as humans.

This is related to what the psychometric tradition addresses through construct validity, convergent validity, and measurement invariance~\cite{cronbach:1955, messick:1987} — formal frameworks for assessing whether a measurement instrument tracks the latent construct it is intended to reflect, though with a focus on groups rather than individuals.

\citet{zajkac:2023} find that omitting background knowledge on a patient from an x-ray labeling task in an attempt to reduce annotator bias paradoxically produces noisier labels, because expert annotators are deprived of information they would rely on in the real clinical task. Thus, omitting features from classification makes outcomes noisy, even when humans are the annotators.

\citet{bengio:2013} surveyed over 200 papers on representation learning in 2013. Almost all of this body of work focuses on representing objects that are either born-digital or digitized, and on compressing them by selecting the most relevant features from a finite set. Only a small fraction considers representations of humans~\cite{athanasiou:2022, zhu:2022, pavlakos:2019}, and these primarily focus on human physiology. The question of whether a feature representation is valid for a given human individual, in the sense of capturing what that individual considers relevant about themselves, remains largely unaddressed.

This fixed feature set creates a circular validation problem: outcomes are evaluated against the same features from which they were constructed, so characteristics outside the set can never challenge the representation. It may appear adequate simply because the population has already been reduced to it.

\paragraph{Representational harm and measurement}

Prior work demonstrates that harms can originate in representation before a model computes a decision outcome. Failures of closed-set representations have been documented in education~\cite{lamont:2021, perdomo:2025}, employment~\cite{bogen:2019, dastin:2019, fanta:2021}, finance~\cite{martinez:2021, andrews:2021, knight:2019}, and public administration~\cite{angwin:2016, mak:2021, constantaras:2023, meaker:2023, oosting:2023, brancolini:2022, geiger:2023}. For example, pandemic-era grading systems relied on teacher-derived rankings that failed to capture individual circumstances~\cite{lamont:2021}.

This concern is formalized in work on measurement and representation. \citet{friedler:2021} distinguish constructs from their observational representations, while \citet{jacobs:2021} and \citet{suresh:2021} show how mismatches between constructs and observations can produce error and inequity. In information retrieval, \citet{tobia:2026, tobia:2026b} similarly show that standardized user models can poorly represent children. These works establish that representation matters, but generally seek better representations for an entire group rather than measuring the fidelity of a representation to each individual.

Work by \citet{lu:2022} and \citet{katzman:2023} emphasizes the importance of self-identification and dynamic, evolving representations, but does not provide methods for comparing these against existing system representations.

\paragraph{Stereotypes and the problem of validation}

The problem is related to the concept of stereotypes. \citeauthor{allport:1954} describes a stereotype as an ``exaggerated belief associated with a category'' used to justify conduct toward that category~\cite[p.191]{allport:1954}. Group-level representations are not necessarily stereotypes in the social sciences sense, but they share the relevant property of reducing heterogeneous individuals to a single limited set of characteristics. Aside from fitting individuals differently well, the validity of representations for the population itself is questionable: if characteristics outside the group-level set are never observed, they cannot challenge the representation. A representation may thus be treated as adequate without ever being tested against the full diversity of characteristics present in the population.

The German Ethics Council has identified ``Risk[s] to the individual posed by statistical stratification''~\cite{deutscherethikrat:2023}. Recently, \citet{hoeltgen:2026} challenge the assumption, prevalent in machine learning practice, that data originates from a pre-existing underlying distribution, arguing that the causal direction is in fact reversed. They call attention to the underexplored role of representational choices, specifically, how the input space and its covariates are defined, as a foundational yet neglected determinant of model behavior. Our work fills this research gap by proposing a validation strategy for representations of individuals that does not assume individuals originate from an underlying distribution.

\paragraph{Human representation validation is a blind spot in fairness research}

Algorithmic fairness research provides extensive methods for evaluating decisions across groups and individuals~\cite{corbett-davies:2023, rudin:2019, corvi:2025}. Many approaches represent individuals as points in a feature space and evaluate distances, similarities, or outcomes within that space~\cite{dwork:2012, yona:2018, kim:2018, gillen:2018, ilvento:2020}. These methods are essential for assessing how decisions are made \emph{from} representations, but generally take the representation itself as given and assume the same degree of faithfulness for each individual.

Individual fairness, introduced by \citet{dwork:2012}, posits that similar individuals should be treated similarly with respect to a given decision task. Rather than grappling with the potentially large set of possible representations of individuals, \citet{dwork:2012} model individuals as single instances in a metric space. Subsequent research has largely focused on defining, learning, or circumventing the need for a universal distance metric between such individuals~\cite{yona:2018, kim:2018, gillen:2018, ilvento:2020}. This line of work assumes that the features supplied to the metric are adequate proxies for the individuals' actual characteristics. Our work questions precisely this assumption: whether the features supplied to the decision process adequately capture the individual being represented. Thus, a decision can satisfy a fairness criterion relative to a feature space while that feature space fails to capture important characteristics of the people within it.

\paragraph{Heterogeneity of a population and Jaccard}

We draw on Paul Jaccard's observation that ``probably no two square metres of vegetation in the whole Alps, possess exactly the same floristic composition''~\cite{jaccard:1912}, projecting square metres to individuals and plants to predictive features. We use the Jaccard Index to quantify representational similarity and individual rarity within a population; formal definitions follow in Section~\ref{sec:measures}.

In what follows, we introduce uniqueness-aware feature engineering, focusing on reducing harm to human individuals through group-level inaccuracies in representation.
\section{A Framework for Uniqueness-Aware Feature Engineering}
\label{sec:framework}

We formalize the individual representation problem and lay out the framework that addresses it. The framework has two parts: a set of entities that define the setting precisely, and a six-step pipeline that operationalizes the framework from data collection through individual-level transparency of representation fit at deployment.

\subsection{Setting}
\label{sec:setting}

\paragraph{Individuals.}
We understand individuals $i \in I$ as real humans with unique and fluctuating traits, experiences, and identities. The individuals in scope are subjects of an algorithmic decision, such as loan-granting or college-admission.

\paragraph{Task.}
The decision task $t$ specifies what the algorithmic system is asked to predict or decide for each individual, such as the probability of credit default, academic success, or criminal re-offense. The task determines which features are potentially relevant and defines what counts as a decision-relevant characteristic in the following.

\paragraph{Reference representations.}
A reference representation $\mathbf{d}_i$ is a per-individual, task-specific characterization of which features are decision-relevant from $i$'s own perspective. For example, reference representations may be self-reported by the individual, provided by a knowledgeable proxy such as a caseworker or teacher. We treat $\mathbf{d}_i$ as a binary feature vector over a possibly open feature space: a feature is either reported (1) or not (0), with no duplicates per individual. Where multiple reference representations exist for one individual, we assume they can be aggregated under deduplication to a single vector that contains each relevant feature exactly once.

Reference representations are a novel construct we introduce in this work. They are not features selected by a data scientist or observed passively in a dataset; they are task-relevant characterizations generated or validated by or for the individual themselves from an internal and subjective perspective. Taking access to reference representations as given is a prerequisite of the framework;
how they are collected is a subject of ongoing work (Section~\ref{future}). 

\paragraph{Pooled population.}
The pooled population $\mathcal{D}$ is the multiset of all features reported across all reference representations in the observed group of individuals during training of the algorithmic decision system:

\[
\mathcal{D} = \bigcup_{i \in I} \mathbf{d}_i.
\]

Some features appear in many reference representations; others appear in only one. This heterogeneity is the central object the framework measures and acts on.

\paragraph{Training population.}
The training population is set of individuals and their reference representations that are observed during the feature engineerig process, including selecting the input representation and training the algorithmic decision system. Before production, the pooled population contains only the reference representations in the training population (and the held-out set).

\paragraph{Input representations.}
An input representation $\mathbf{x}$ is a fixed feature set used by an algorithmic decision system to represent every individual.
It is the same for all individuals and is derived from $\mathcal{D}$ through feature selection.
The individual representation problem arises precisely from the conflict between the population-level $\mathbf{x}$ and the individual $\mathbf{d}_i$: There is a conflict between the coverage of all individually decision relevant features in a heterogeneous population and the population-level useful features to compute a statistically grounded algorithmic decision with.

\subsection{The Uniqueness-Aware Feature Engineering Pipeline}
\label{sec:pipeline}

We operationalize the framework as a six-step pipeline, illustrated in
Figure~\ref{fig:uniqueness-aware-framework}.
Steps~1--2 characterize the population before any feature is selected.
Steps~3--4 govern feature selection and model training.
Steps~5--6 fulfil transparency obligations at deployment and in production.

\paragraph{Step~1: Collect reference representations.}
Gather $\mathbf{d}_i$ for all individuals in the training population.
The collection process should be task-specific: individuals (or knowledgeable proxies like case workers) document what they consider decision-relevant for for this individual and the particular task at hand, despite of what is observeable in an entire group. We assume the resulting reference representations are complete and truthful to the best of each individual's knowledge; limitations arising from imperfect self-knowledge are discussed in Section~\ref{challenges}.

\paragraph{Step~2: Characterize population heterogeneity.}
Compute the measures introduced in Section~\ref{sec:measures}---full coverage probability, similarity (here evaluated pairwise across the population), and rarity---for every individual in the training population.
This step answers the question: How diverse is the pooled population, and are minority subgroups systematically rarer?
A population in which most individuals report nearly the same features is tractable for uniform feature selection; a population in which individuals report highly idiosyncratic feature sets signals that any fixed input representation will leave many poorly covered, and alternative strategies to algorithmic decisions, such as human oversight, or human-only decisions, should be considered.

\paragraph{Step~3: Select features by prevalence.}
Choose a prevalence threshold $\tau \in [0,1]$ and retain in the input representation $\mathbf{x}$ all features whose prevalence $p(f)$, i.e. the fraction of individuals who report it, in the pooled population meets or exceeds $\tau$:

\[
\mathbf{x}(\tau) = \bigl\{f \in \mathcal{D} \mid p(f) \geq \tau\bigr\}.
\]

Features below $\tau$ are excluded on the grounds that too few reference representations support reliable estimation of their predictive effect. The choice of $\tau$ is calibrated in Step~4.

\paragraph{Step~4: Calibrate the coverage--accuracy optimum.}
Raising $\tau$ prunes rare features, which can improve predictive accuracy up to the point where the group is underspecified but also reduces the fraction of each individual's reference representation that $\mathbf{x}$ covers. Low coverage leads to poor representational fit of affected individuals.
This optimum is not incidental: it is a structural consequence of the individual representation problem, and it does not only fall on individuals but aggregates on minority subgroups whose relevant features are precisely the ones most likely to fall below any positive threshold.

We make this optimization problem explicit by sweeping $\tau$ and computing, for each candidate threshold, mean individual coverage and held-out predictive accuracy (cf. Section~\ref{sec:measures}). A minimum coverage floor is set (e.g. by the deploying institutions and society). $\tau$ is chosen so that accuracy is maximized within this admissible region above the coverage floor. Additionally reporting worst-group coverage makes baldy represented subgroups visible, which is a direct consequence of the threshold choice.

\paragraph{Step~5: Publish a structured representation report.}
Before deployment, publish a structured account of the chosen input representation: which features are included and which are excluded, the full calibration curve (coverage and accuracy as a function of $\tau$), the chosen threshold and its justification, and the group-level coverage gaps at that threshold. This report is the transparency artefact that enables external auditing, regulatory review, and institutional accountability. It makes explicit what any feature selection decision implies for different individuals and groups: Information that is currently absorbed silently into model training with no external record.

\paragraph{Step~6: Inform each individual of their fit.}
During production (at decision time), compute the individual coverage (Section 4) for each new individual against the deployed input representation and potentially assign them coverage classes for simpler communication — e.g., poorly, partially, or well represented — defined by percentile thresholds of that score in the training population.
The individual fit report also identifies any features in $\mathbf{d}_i$ that are absent from the training population (\emph{never-seen features}): such features systematically lower the individual coverage and if repeatedly reported during production may signal that the deployed representation may need recalibration to include emerging characteristics not present when the system was trained.

\section{Measures}
\label{sec:measures}

We introduce a set of measures that together assess population heterogeneity, guide feature selection, and quantify individual and system-wide representation fit. Prevalence and rarity operate at the feature level and characterize the pooled population, full coverage probability characterizes how distinctive each individual reference representation is within that population, similarity and coverage measure the fit between a chosen input representation and an individual reference representation. Aggregates of these measures to system-level and subgroup-level quantities are used during calibration and at deployment.

\subsection{Prevalence and Rarity}
\label{sec:prevalence}

Let $k(f)$ denote the number of reference representations in the pooled population that contain
feature $f$, and let $N = |I|$ be the total number of individuals.
The \emph{prevalence} of feature $f$ is the fraction of individuals who report it:

\[
p(f) = \frac{k(f)}{N}.
\]

The \emph{rarity} of feature $f$ is the complement of its prevalence---the fraction of individuals
who do \emph{not} report it:

\[
\text{Rare}(f) = 1 - p(f) = 1 - \frac{k(f)}{N}.
\]

A feature with $\text{Rare}(f) \approx 1$ appears in almost no reference representations;
a feature with $\text{Rare}(f) \approx 0$ is reported by nearly everyone.

The rarity of an individual's reference representation $\mathbf{d}_i$ summarizes how uncommon
their reported features are in the pooled population.
We use the median rarity over the individual's reported features:

\[
\text{Rare}(\mathbf{d}_i) = \underset{f \in \mathbf{d}_i}{\operatorname{median}} \; \text{Rare}(f).
\]

Prevalence drives feature selection (Step~3): features with $p(f) \geq \tau$ are retained in the
input representation $\mathbf{x}(\tau)$; features below the threshold are excluded.
Rarity characterizes which individuals are most affected by that exclusion.

\subsection{Full Coverage Probability}
\label{sec:collision}

Full coverage probability $P_i$ is a special case of the multivariate hypergeometric distribution, adapted for our binary setting: each feature in the pooled population is either reported by an individual or it is not. This corresponds to an urn model in which each ball can be drawn at most once. Hence, $P_i$ answers ``Given the pooled population, how likely is it that a randomly drawn feature set  exactly matches individual $i$'s reference representation $\mathbf{d}_i$?''
A low $P_i$ indicates that $i$'s feature combination is highly distinctive and unlikely to arise by chance from the population distribution.

Sampling without replacement, the probability that the features actively reported by individual $i$ are drawn is:

\[
P_i = \frac{\displaystyle\prod_{j:\mathbf{d}_i[j]=1} a_j}
           {\displaystyle\binom{\sum_{j=1}^{|\mathbf{d}_i|} a_j}{||\mathbf{d}_i||_1}}
\]

where $a_j$ is the count how often a specific feature at position j in $i$'s reference representation appears in the pooled population given that it was actively reported by individual $i$ (i.e., where $\mathbf{d}_i[j] = 1$).

Because $P_i$ reflects both the number of reported features and how common those features are, it serves as a compact measure of how easy or difficult it is to match any specific individual with a uniform input representation. In heterogeneous populations, $P_i$ is typically vanishingly small for all individuals, and is best interpreted on a $\log_{10}$ scale or relative to the population distribution of $\log_{10} P_i$.

The most demanding scenario for feature engineering arises when $P_\text{system}$ is low and the distribution of $P_i$ is (approximately) identical. In this case, no single input representation is likely to adequately cover most individuals, and there is no exploitable variation in individual representability to guide the decision on which features should compose $\mathbf{x}$. This configuration of the full coverage probability distribution is a signal that a single pooled feature set may be fundamentally insufficient, and that alternative strategies---individualized decisions, or explicit flagging of poorly covered instances at deployment---should be considered before proceeding with a uniform feature selection approach.

\subsection{Coverage and Similarity}
\label{sec:similarity}

The coverage of input representation $\mathbf{x}$ for individual $i$'s reference representation $\mathbf{d}_i$ is their Jaccard index:

\[
J_i(\mathbf{x}, \mathbf{d}_i) = \frac{|\mathbf{x} \cap \mathbf{d}_i|}{|\mathbf{x} \cup \mathbf{d}_i|}.
\]

$J_i = 1$ means the input representation is a perfect match for $i$'s reference representation; $J_i = 0$ means no overlap. We suggest to implement thresholds (e.g. as percentile fit classes as implemented in~\ref{sec:prototype}) of poorly represented individuals that in order to flag those for human review rather than automated processing.

Three system-level aggregates of $J_i$ are used during calibration and auditing:

\paragraph{Population Similarity ($J_\text{pop}$).}
The mean pairwise Jaccard index of reference representations in the pooled population:

\[
J_\text{pop} = \frac{2}{N(N-1)} \sum_{i < i'} J(\mathbf{d}_i, \mathbf{d}_{i'}).
\]

$J_\text{pop}$ measures how similar individuals are to one another in their reference representations, independently of any chosen input representation. A low $J_\text{pop}$ indicates high population heterogeneity, which signals that uniform feature selection will inevitably leave many individuals partially covered.

\paragraph{System Coverage ($J_\text{system}$).}
The mean coverage across all individuals in the reference population at a given threshold $\tau$:

\[
J_\text{system}(\tau) = \frac{1}{N} \sum_{i=1}^{N} J_i\bigl(\mathbf{x}(\tau), \mathbf{d}_i\bigr).
\]

$J_\text{system}$ is the primary quantity used to calibrate $\tau$ (Step~4): it measures what fraction of each individual's reference features the chosen input representation captures on average.

\paragraph{Worst-group coverage ($J_\text{worst}$).}
The minimum of $J_\text{system}$ computed separately for each subgroup $g \subseteq I$:

\[
J_\text{worst}(\tau) = \min_{g} \; \frac{1}{|g|} \sum_{i \in g} J_i\bigl(\mathbf{x}(\tau), \mathbf{d}_i\bigr).
\]

$J_\text{worst}$ surfaces subgroup-level representational disparities that are averaged away in $J_\text{system}$, and is published as part of the structured representation report (Step~5).

These measures decompose the individual representation problem along four axes: which features the population can reliably support (prevalence), how unusual an individual’s features are within that population (rarity), what a specific chosen input representation actually achieves for each individual (coverage), and how likely it is to find an input representation with perfect coverage given the population structure (full coverage probability). Together they cover population characterization and individual assessment, with prevalence and rarity grounding feature selection for the input representation in the training population's specific heterogeneity, and full coverage probability and coverage and similarity providing the before-and-after view of representational fit for each individual and deploying institution.

\section{Prototype on Synthetic Data}
\label{sec:prototype}

We illustrate the six-step pipeline on a controlled synthetic dataset in which ground truth is fully known. The goal is not to validate the measures on real individuals---that requires empirical work discussed in Section~\ref{future}---but to trace the pipeline end to end and make its outputs concrete.~\footnote{All code and data is available at \url{https://anonymous.4open.science/r/uniqueness-aware-feature-engineering/}.}

\subsection{Synthetic population}

\paragraph{Data generation (Step 1).}
We draw a population of $N_\text{ref} = 400$ individuals whose reference representations are binary feature vectors over $F = 570$ reported (1) or not reported (0) possible features. The feature space is partitioned into three kinds: \emph{Shared features} (150) are reported by most individuals regardless of subgroup; \emph{Subgroup features} (120) are specific to one of four subgroups A--D, with population shares 60\,\%, 25\,\%, 10\,\%, and 5\,\%; and \emph{Idiosyncratic features} (300) are sampled from a long-tailed distribution; many are reported by only one or two individuals.

Each individual's feature count follows the same long-tailed distribution (median 29 features). A binary outcome $y_i$ is assigned to each individual. An additional $N_\text{new} = 100$ individuals are held back as \emph{new arrivals} to simulate representation reports in production in Phase~III; arrivals also report brand-new features---feature keys that no training individual reported---at an average rate of 1.5 per arrival.

\subsection{Phase I: assessing population heterogeneity (Step 2)}

\paragraph{Population-level measures.}
Table~\ref{tab:prototype-results}~(a)-(Phase I) summarizes the pooled population after computing the measures for every individual in the training data. Note that \emph{every individual has a unique feature vector}: no two individuals in the training population share exactly the same set of reported features. Full coverage probabilities quantify how unlikely it is that a randomly drawn input representation would exactly match an individual's reference representation. Because every individual has a unique feature vector, these probabilities are vanishingly small for everyone: median $\log_{10} P_i$ ranges from $-34.9$ for the majority subgroup~A to $-41.5$ for subgroup~C (corresponding to probabilities of roughly $10^{-35}$ to $10^{-42}$). The more negative values for minority subgroups reflect that their subgroup-specific features appear in only a small fraction of the pooled population, making their exact feature combinations rarer still. Rarity scores are uniformly high ($\text{Rare}(d_i) = 85.2\,\%$), meaning the typical individual reports features that fewer than 15\,\% of the population also report. Population-level Jaccard similarity ($J_\text{pop}$, the mean pairwise Jaccard index across all reference representations) is $0.132$: any two individuals share, on average, only 13\,\% of their reported features.

\begin{figure}[t]
  \centering
  \includegraphics[width=\linewidth]{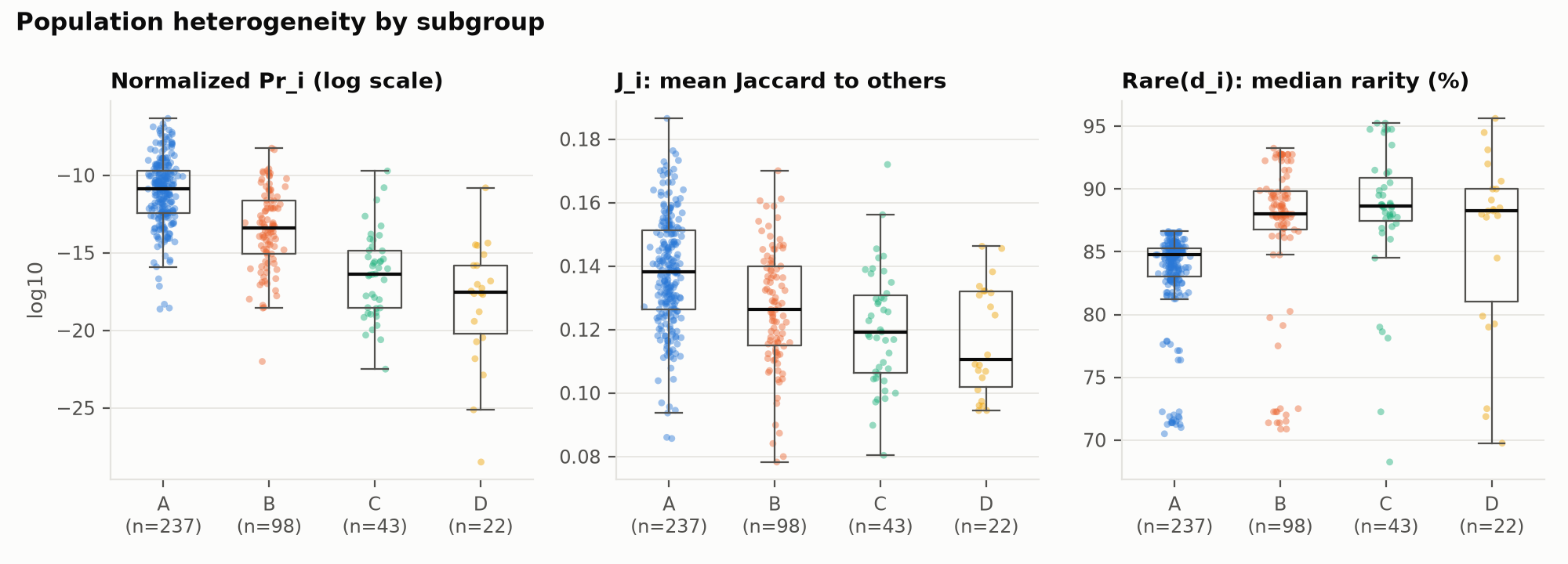}
  \caption{Distribution of the measures by subgroup (Phase~I).
    Full coverage probabilities are shown on a $\log_{10}$ scale after normalization.
    Minority subgroups (C, D) exhibit lower full coverag probabilities and higher rarity scores,
    reflecting their subgroup-specific and idiosyncratic feature sets.}
  \label{fig:measures}
\end{figure}

\paragraph{Minority subgroups are systematically rarer.}
Subgroup D's individuals have a median rarity of 88.3\,\% and a median $\log_{10} P_i$ of $-41.0$, compared with 84.8\,\% and $-34.9$ for the majority subgroup A (Table~\ref{tab:population}). This reflects the design of the synthetic population---subgroup-specific features are, by construction, uncommon across the full pooled population---but the same dynamic is expected in real-world populations where minority characteristics are underrepresented in data collection. Figure~\ref{fig:measures} visualizes the measure distributions by subgroup.

\subsection{Phase II: feature selection and calibration (Steps 3--4)}

\paragraph{Prevalence threshold sweep.}
 We select an input representation $\mathbf{x}$ by prevalence, retaining all features that meet or exceed a prevalence-threshold $\tau$ in the pooled population. Prevalence-based selection represents the best-case scenario for representational coverage: because $\mathbf{x}$ is grounded in what the training population actually reports, most individuals are well-served by it. An input representation chosen without reference to the observed population---in the worst case, one containing features nobody would ever report---would yield poor coverage for nearly everyone, making prevalence the natural baseline against which more targeted selection strategies should be measured.
For each $\tau$, we compute via five-fold cross-validation: (i)~$J_\text{system}$, the mean share of each held-out individual's reference features that the system retains---our measure of representational coverage---and (ii)~held-out predictive accuracy using logistic regression.

\begin{figure}[t]
  \centering
  \includegraphics[width=\linewidth]{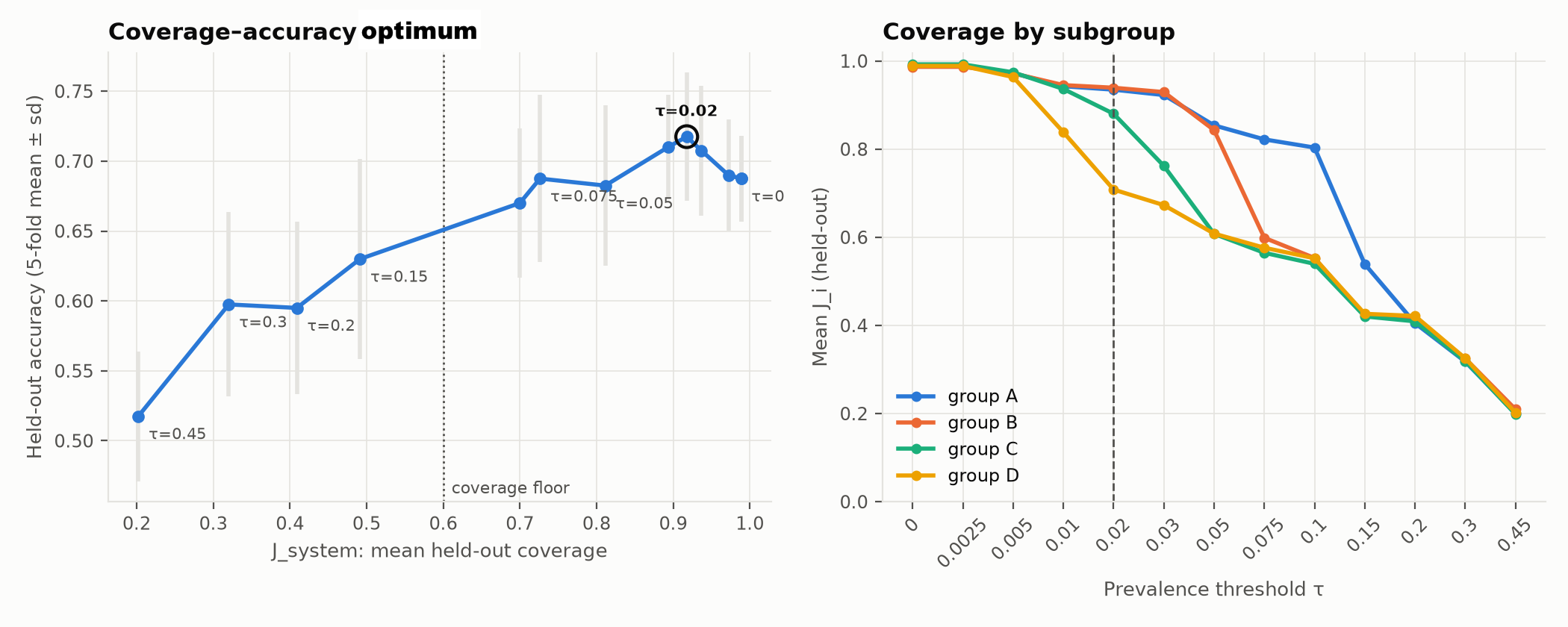}
  \caption{Coverage--accuracy optimization as a function of the prevalence threshold $\tau$
    (Phase~II).
    The shaded region marks $J_\text{system} < 0.60$, below the minimum coverage requirement.
    The chosen threshold ($\tau = 0.02$, dotted line) maximises held-out accuracy within the
    admissible region.
    Per-group $J_\text{system}$ curves show that coverage degrades fastest for minority subgroups
    as $\tau$ increases.}
  \label{fig:calibration}
\end{figure}

Table~\ref{tab:calibration} and Figure~\ref{fig:calibration} show the resulting optimization function. A low $\tau$ keeps almost the entire feature space ($\tau = 0$: 539 features, $J_\text{system} = 0.989$) but yields accuracy of only $0.688$. Raising $\tau$ prunes rare features, reducing coverage while initially increasing accuracy. The chosen threshold $\tau = 0.02$ selects 236 features, achieves $J_\text{system} = 0.918$ and accuracy $0.717 \pm 0.046$, and is the lowest threshold that keeps $J_\text{system} \geq 0.60$---our minimum coverage requirement. Above $\tau = 0.10$ coverage and accuracy decline together: the representation no longer captures enough signal to be predictive.

\paragraph{Subgroup coverage gaps.}
Even at the chosen threshold, representational coverage is unevenly distributed (Table~\ref{tab:prototype-results})~(a)-(Phase II). Majority subgroup A keeps 93.5\,\% of its features on average; minority subgroup D keeps only 70.8\,\%. This gap arises because subgroup-specific features of smaller groups fall below the prevalence threshold: a feature reported by 5\,\% of the population meets $\tau = 0.02$ in expectation, but when the subgroup itself constitutes only 5\,\% of individuals, most of its distinctive features fall below $\tau$. The worst-group coverage $J_\text{worst}$ drops from 0.987 at $\tau = 0$ to 0.708 at the chosen threshold---a direct, measurable consequence of applying a minimum prevalence requirement.

\subsection{Phase III: deployment (Steps 5--6)}

\paragraph{Representation report (Step 5).}
The deployed system uses 241 of the 570 features from the pooled population: 149 of 150 shared features, 91 of 120 subgroup features, and 1 of 300 idiosyncratic features. The single retained idiosyncratic feature meets the prevalence threshold because it happens to be reported by a sufficient fraction of the training population; the remaining 299 idiosyncratic features are pruned. The structured representation report packages these facts alongside the calibration summary and the per-group coverage gap for public release.

\paragraph{Individual fit scoring at arrival time (Step 6).}
Each of the 100 new arrivals is scored against the deployed system using their reported feature names. Table~\ref{tab:prototype-results}~(a)-Phase III summarizes the results. Of the 100 arrivals, 50 are \emph{well represented} ($J_i \geq 0.90$), 48 are \emph{partially represented} ($0.75 \leq J_i < 0.90$), and 2 are \emph{poorly represented} ($J_i < 0.75$). All poorly-represented individuals belong to subgroup D.

\begin{table*}[t]
\caption{Prototype evaluation results across all phases (Section~\ref{sec:prototype}).
  \textit{(a)}~Population-level uniqueness measures (Phase~I), per group coverage and accuracy at $\tau=0.02$ (Phase~II), and per group coverage classes for 100 new arrivals (Phase~III).
  \textit{(b)}~Coverage--accuracy optimization across prevalence thresholds (Phase~II).
  \textit{(c)}~Effect of never-seen features on fit score and coverage classes (Phase~III).}
\label{tab:prototype-results}
\smallskip
\begin{minipage}[t]{\linewidth}
  \centering
  \small(a)
  \label{tab:population}\\[0.5ex]
  \begin{tabular}{lrrrrrrrrrr}
  \toprule
  \multicolumn{6}{c}{Phase~I} & \multicolumn{2}{c}{Phase~II} & \multicolumn{3}{c}{Phase~III} \\
  \cmidrule(r){1-6}\cmidrule(lr){7-8}\cmidrule(l){9-11}
  Group & $n$ & Features & $\log_{10} P_i$ & $J_\text{pop}$ & $\text{Rare}(d_i)$\,\% & $J_\text{system}$ & Accuracy & Well repr. & Partially repr. & Poorly repr.\\
  \midrule
  A & 237 & 29 & $-34.9$ & 0.138 & 84.8 & 0.935 & 0.734 & 32 & 31 & 0 \\
  B &  98 & 29 & $-37.1$ & 0.127 & 88.0 & 0.939 & 0.633 &  17 & 10 & 0\\
  C &  43 & 31 & $-41.5$ & 0.120 & 88.6 & 0.880 & 0.814 &   1 &  6 & 0 \\
  D &  22 & 28 & $-41.0$ & 0.117 & 88.3 & 0.708 & 0.727 &   0 &  1 & 2 \\
  \midrule
  All & $\sum$ 400 & Avg 29 & $-35.7$ & 0.132 & 85.2 & 0.866 & 0.727 & $\sum$ 50 & 48 & 2 \\
  \bottomrule
  \end{tabular}
\end{minipage}
\par\medskip
\begin{minipage}[t]{0.30\linewidth}
  \small(b)
  \label{tab:calibration}\\[0.5ex]
  \centering
  \begin{tabular}{rrrrr}
  \toprule
  $\tau$ & Feat. & $J_\text{system}$ & $J_\text{worst}$ & Acc. \\
  \midrule
  0.000 & 539 & 0.989 & 0.987 & 0.688 \\
  \textbf{0.020} & \textbf{236} & \textbf{0.918} & \textbf{0.708} & \textbf{0.717} \\
  0.100 &  95 & 0.700 & 0.540 & 0.670 \\
  0.450 &  10 & 0.202 & 0.199 & 0.518 \\
  \bottomrule
  \end{tabular}
\end{minipage}
\hspace{1em}
\begin{minipage}[t]{0.61\linewidth}
  \small(c) 
  \label{tab:unseen}\\[0.5ex]
  \centering
  \begin{tabular}{lrrrrrr}
  \toprule
  Unseen & $n$ & $J_i$ (mean) & Percentile (med.) & Well & Partial & Poor \\
  \midrule
  0      & 21 & 0.938 & 67.5 & 18 &  3 & 0 \\
  1      & 33 & 0.907 & 34.1 & 22 & 11 & 0 \\
  2      & 31 & 0.867 & 15.6 & 10 & 19 & 2 \\
  3+     & 15 & 0.838 &  9.4 &  0 & 15 & 0 \\
  \bottomrule
  \end{tabular}
\end{minipage}
\end{table*}

\paragraph{Never-seen features as a signal.}
Seventy-nine of the 100 arrivals report at least one feature that no individual in the reference population reported. Table~\ref{tab:prototype-results}~(c) shows that the presence of never-seen features consistently lowers the fit score ($J_i$): arrivals reporting zero unseen features have a mean fit score of 0.938 and a median percentile of 67.5; those reporting three or more unseen features have a mean fit score of 0.838 and a median percentile of 9.4, placing them in the bottom decile of the pooled population's coverage distribution. 85 distinct never-seen features appear across the 79 affected arrivals; several are reported by multiple arrivals (e.g. two keys each appear in four arrival reports), suggesting that some represent emergent subgroup characteristics the training data did not yet capture. These features are flagged as candidates for the next recalibration pass.

\paragraph{Examplary individual fit report.}
A new arrival receiving a low fit score illustrates Step~6. \emph{Individual new\_074} (subgroup D) is poorly represented: $J_i = 0.69$ at the 2nd percentile of the reference population. The system keeps 20 of 29 reported features; the nine omitted features include six subgroup-specific features, one idiosyncratic feature, and two features the system has never seen. Their rarity score is $\text{Rare}(d_i) = 87.3\,\%$ and $\log_{10} P_i = -43.4$. 
This individual is flagged by the pipeline as candidate for human review rather than automated processing.~\footnote{These are only examples, what could be done with low scores, also the precise thresholds on when to abstain from algorithmic decisions must be decided by domain experts in the decision tasks and society.}

\paragraph{Summary of prototype findings}
The prototype confirms three structural properties of the framework: uniqueness is universal — every individual in the training population has a distinct feature vector and vanishingly small full coverage probability — minority subgroups are systematically disadvantaged at any positive prevalence threshold, and the coverage–accuracy optimization is empirically non-trivial, with the threshold that maximizes accuracy entailing measurable representational loss concentrated in the least-covered groups. Never-seen features additionally serve as a recalibration signal: their systematic degradation of individual fit scores provides an operational mechanism for detecting when the deployed representation has drifted from the population it serves.






\section{Discussion}

Heterogeneity within a population is not merely a statistical property to be acknowledged and set aside; it is a fundamental determinant of which input representation best serves that population. Feature engineering must therefore be grounded in an understanding of the population a model is trained on, and the variation of individuals with respect to any chosen representation must be measured and communicated. This assessment cannot be performed once at training time and assumed to hold indefinitely: it must extend to individuals newly entering the system in production.

Our feature engineering framework and associated prototype build on the notion that uniqueness of the represented individuals is the norm, not the exception. It does not overcome harm through misrepresentation, but in contrast to traditional feature engineering strategies, makes this harm visible and quantifies it. Trustworthy algorithmic decision-making requires structures that communicate to affected individuals and deploying institutions how appropriate it is to apply a formalized decision to any specific person, and that allow the public to assess for how whom algorithmic treatment could be appropriate and for how whom human oversight is warranted.

\paragraph{When is uniqueness-aware feature engineering necessary?}
The approach is most clearly warranted when the subjects of an algorithmic decision are human individuals: they are unique, they hold an internal perspective on what is predictive for them, and they can be structurally harmed when that perspective is ignored. The boundary is less sharp for non-human subjects. At first glance, an artifact like a vehicle lacks moral standing and is not exposed to structural bias. On closer inspection, however, uniqueness resurfaces: individual driving behavior and usage patterns may carry high predictive value yet remain invisible to conventional feature engineering, and the stakes of misclassification can be severe. A principled guideline delineating when uniqueness-aware feature engineering is warranted---and when the added complexity is not---remains an open task.

\paragraph{Benefits of uniqueness-aware feature engineering}
At the individual and group level, the framework provides a principled basis for quantifying how well an input representation captures what is genuinely predictive for a specific person or group, enabling transparency grounded in features the individuals themselves would recognize as relevant. At the system level, aggregate measures allow institutions and auditors to assess not merely whether a model is accurate on average, but whether its accuracy is fairly distributed across the population it governs. For individuals and groups whose features fall outside the chosen feature set, the framework provides a mechanism to flag cases for human review rather than allowing an unjustified automated decision to stand unchallenged.

\paragraph{Challenges of uniqueness-aware feature engineering}
\label{challenges}
The framework assumes that each person's reference representation is complete and correct---that it accurately captures all characteristics predictive of the decision outcome for that individual. In reality, individuals may forget relevant features, may not know which of their characteristics are predictive, or may misjudge their own relevance. We believe this does not undermine the framework's value: if an individual regards their reference representation as adequate, our measures can quantify their input representation's adequacy from their perspective. At the system level, aggregated subjective assessments---even if individually skewed---carry a meaningful signal about a system's faithfulness.

A deeper structural concern is different access to realistic reference representations. For example, people differ in their ability to articulate what is decision-relevant for them, and this variation may systematically disadvantage individuals or groups. The framework must be carefully adapted to mitigate this risk.

Practical challenges center on constructing reference representations. It is not clear which modality best conveys decision-relevant information for a given task, nor what should be extracted---feature keys, values, or both---and how similar types of information should be grouped. Each extraction method introduces its own systematic distortions, which compound across the pipeline. Methods that make these distortions legible to researchers, institutions, and affected individuals are as urgently needed as the extraction methods themselves.

\subsection{Future Work}
\label{future}

Before the framework can be applied at scale, empirical and methodological groundwork is needed to establish how reference representations are created---who produces them, in what medium, in what format---and how the resulting distributional properties vary across domains and tasks. Reliable conversion of raw representations into structured feature vectors is a prerequisite. We regard natural language texts, such as task-specific self-descriptions of individuals, processed with NLP methods, as the most promising direction. A specific challenge here is semantic equivalence: two people sharing a relevant feature may describe it in entirely different words, and mapping such surface variation onto a common feature space requires robust paraphrase detection and semantic normalization~\cite{jaccard:1912}.

An important extension is to ask whether representational adequacy can degrade over time. Our measures offer a detection mechanism: monotonically declining similarity scores or full coverage probabilities for individuals entering the system in production suggest the representation is drifting out of alignment with lived reality. Repeated reporting of the same features that have not been observed in the training population but accumulate in production may reveal emergent subgroups absent from the training data and be also a signal for retraining the decision system.
\section{Conclusion}
This work addresses the individual representation problem: no group-level input representation can faithfully capture all individuals, and the resulting gap between what group-level features the system needs and what is decision-relevant in an individual level carries substantial consequences for the justifiability of resulting decisions. We address it with a framework and prototype operating in two layers: the first diagnoses population heterogeneity, revealing where the difficulty of the individual representation problem for the training population; the second acts on this diagnosis by choosing the input representation with optimal representational coverage and accuracy. Operationalized as a six-step pipeline, the framework grants individuals agency over what counts as decision-relevant, gives institutions a quantitative basis for accountability through a publishable representation fit report, and upholds transparency by informing each person of their representational fit at deployment.

\newpage
\section{Endmatter}
\subsection{Generative AI Usage Statement}
The authors used the following generative AI tools during the preparation of this manuscript:

Claude Sonnet 4.6 (Anthropic, accessed via Claude desktop application) was used for English language editing and style improvements, brainstorming paper titles, developing names for proposed concepts (e.g., the “individual representation problem”), generating a mock peer review to anticipate reviewer feedback, and suggesting ways to shorten and tighten the manuscript.

Claude Code with Opus 5.5 (Anthropic, accessed via Visual Studio Code) was used to implement the final prototype based on a code draft and specification of the 6-step pipeline written by the authors.

Gemini 3.5 Flash (Google, accessed via Google Colab) was used to assist with coding the initial draft notebook.

The authors retain full responsibility for the originality and integrity of all content, including the correctness of all claims, references, and citations. All AI-generated suggestions were reviewed, verified, and revised by the authors prior to inclusion.

\subsection{Ethical Considerations} 

The proposed approach introduces several potential risks. First, collecting and processing real reference representations (as opposed to abstract, synthetic ones in the paper) raises concerns about privacy and informational self-determination, as individuals may disclose sensitive attributes that could be misused. In the event of data breaches, such information could be exploited to train systems for harmful or discriminatory purposes.

Second, same as input representations, reference representations may lead to unequal representation fit: individuals differ in their ability of expression, which may systematically disadvantage certain groups. This can result in biased or inaccurate measurements that affect downstream decisions and fit reports.

Finally, improving the apparent trustworthiness and interpretability of decision systems may create a false sense of reliability. Systems that appear transparent or aligned may be less critically scrutinized, making harmful behavior harder to detect and potentially increasing their impact.

\subsection{Positionality Statement}

Our perspective is shaped by a Western, privileged academic context with a background in Computer Science and Arts. Accordingly, the problems addressed in this work reflect concerns emerging in relatively well-resourced settings and should not be prioritized over more immediate threats to safety, rights, and well-being in less privileged contexts.

We take an individual-centered view on fairness, focusing on alignment between generalized system representations and individuals’ unique reality. This choice reflects awareness of the limits of our own perspective---we cannot assume knowledge of all relevant or marginalized groups. 

Our work is further informed by a background that emphasizes the uniqueness and irreducibility of human experience. Motivated by concerns about the simplification inherent in computational systems, we aim to identify and question reductions of complex human realities, while recognizing that our perspective is partial and that broader impact requires engagement beyond it.

\bibliographystyle{ACM-Reference-Format}
\bibliography{arxiv26-uniqueness-aware-feature-engineering-lit}

\end{document}